\documentclass[copyright,creativecommons]{eptcs}
\providecommand{\event}{M.H. ter Beek and M. Loreti (Eds.): Workshop on FORmal methods for the quantitative Evaluation of Collective Adaptive SysTems (FORECAST'16).} % Name of the event you are submitting to
\usepackage{graphicx}   
\usepackage[dvipsnames]{xcolor}
\usepackage{fancyvrb}   
\usepackage{subfigure}   

\title{Resiliency with Aggregate Computing:\\ State of the Art and Roadmap}
\author{Mirko Viroli
\institute{Alma Mater Studiorum -- Universit{\`a} di Bologna\\ Cesena, Italy}
\email{mirko.viroli@unibo.it}
\and
Jacob Beal
\institute{Raytheon BBN Technologies\\
Cambridge, MA, USA}
\email{jakebeal@bbn.com}
}
\def\titlerunning{Resiliency with Aggregate Computing: State of the Art and Roadmap}
\def\authorrunning{Mirko Viroli \& Jacob Beal}
\begin{document}
\maketitle

\begin{abstract}
One of the difficulties in developing collective adaptive systems is the challenge of simultaneously engineering both the desired resilient behaviour of the collective and the details of its implementation on individual devices.
Aggregate computing simplifies this problem by separating these aspects into different layers of abstraction by means of a unifying notion of computational field and a functional computational model.
We review the state of the art in aggregate computing, discuss the various resiliency properties it supports, and develop a roadmap of foundational problems still needing to be addressed in the continued development of this emerging discipline.
\end{abstract}

% %% Space-time Computations:
% Meta-command:

% Code highlighting
\newcommand{\il}[1]{{\it \textcolor{gray}{// #1}}} % inline comment
\newcommand{\pr}[1]{\textcolor{OliveGreen}{#1}} % primitives
\newcommand{\km}[1]{\textcolor{purple}{#1}} % key mechanism primitives
\newcommand{\ex}[1]{\textcolor{blue}{#1}} % external imported Java values
\newcommand{\fc}[1]{\textcolor{blue}{#1}} % field calculus calls
\newcommand{\fn}[1]{\textcolor{blue}{#1}} % building block / function calls
\newcommand{\op}[1]{\fn{#1}} % building block / function calls
\newcommand{\vb}[1]{\textcolor{OliveGreen}{#1}} % variables
\newcommand{\str}[1]{\textcolor{darkgray}{#1}} % strings

% Syntax:
% Field calculus:
\newcommand{\comment}[1]{\hfill\footnotesize{\textcolor{gray}{\text{;; {#1}}}}}
\newcommand{\BNFcce}{{\bf ::=}}
\newcommand{\BNFmid}{\;\bigr\rvert\;}
\newcommand{\oname}{\texttt{b}}
\newcommand{\fname}{\texttt{f}}
\newcommand{\lit}{\texttt{l}}
\newcommand{\var}{\texttt{x}}
\newcommand{\expr}{\texttt{e}}
\newcommand{\we}{\texttt{w}} % variable or (local) value
\newcommand{\PROGRAM}{\texttt{P}}
\newcommand{\FUNCTION}{\texttt{F}}
\newcommand{\e}{\texttt{e}}
\newcommand{\s}{\texttt{s}}
% Keywords
\newcommand{\defK}{\texttt{def}}

% Eventually Approximable:
\newcommand{\senseK}{\texttt{sense}}
\newcommand{\Boolean}{\spaceof{B}}
\newcommand{\Integer}{\spaceof{Z}}
\newcommand{\Real}{\spaceof{R}}
\newcommand{\local}{\texttt{o}}
\newcommand{\comparator}{\texttt{<}}
\newcommand{\cmath}{\texttt{m}}
\newcommand{\mux}{\texttt{mux}}
\newcommand{\OneOp}{\texttt{GPI}} % gradient-path-integral
\newcommand{\boundary}{\mathcal{B}}
\newcommand{\calculus}{GPI-calculus{}}
\newcommand{\validdomain}[1]{#1^{-1}(\spaceof{\datavalue}-\boundary)}

\newcommand{\ltrue}{\textit{true}}
\newcommand{\lfalse}{\textit{false}}

\newcommand{\bral}{\textrm{{\tt {\char '173}}}\,}
\newcommand{\brar}{\textrm{{\tt {\char '175}}}}
\newcommand{\asgK}{~\texttt{=}~}
\newcommand{\letK}{\texttt{let}~}
\newcommand{\tupK}[1]{\texttt{[}#1\texttt{]}}
\newcommand{\lambdaK}[2]{\texttt{(}#1\texttt{)->}#2}
\newcommand{\bodyK}[1]{\bral\! #1\!\brar}
\newcommand{\dotK}{\texttt{.}}
\newcommand{\applyK}{\texttt{apply}}
\newcommand{\mname}{\texttt{m}}
\newcommand{\aname}{\texttt{\#a}}
\newcommand{\repK}[3]{\texttt{\fc{rep}(#1\fc{<-}#2)}#3}
\newcommand{\ifK}[3]{\texttt{\fc{if}}(#1)#2\,\texttt{\fc{else}}\,#3}
\newcommand{\muxK}[3]{\texttt{\fc{mux}}(#1)#2\,\texttt{\fc{else}}\,#3}
\newcommand{\nbrK}[1]{\texttt{\fc{nbr}}#1}

\section{Introduction}

The environment in which we all live, work, and play is increasingly saturated with computational devices, and those devices are increasingly linked with one another, with the physical environment, application services, and humans.
The problems and applications of this emerging computational environment are being addressed in a wide variety of different areas, including such areas as smart cities, intelligent transportation systems, personalized health care, and the Internet of Things.
A common problem in all such diverse scenarios is to tractably engineer safe, reliable, and maintainable collective behaviours in a complex open environment comprising many devices and scales of operation.

Aggregate computing is an approach to these problems based on the recognition that many collective applications are most naturally specified in terms of aggregate properties, rather than the behaviour of individual devices.
For example, 
a crowd safety service needs to know the density and distribution of people through the environment, not the location of individuals, 
and users of a bike-sharing system do not typically care which bicycle or station they use as long as one is readily available nearby.
Building on the natural expression of such properties in terms of collections of values spread over regions of space, called \emph{computational fields} \cite{tota,SpatialIGI2013},
aggregate computing factors the challenging problems of building collective adaptive systems into several abstraction layers, each of which can be engineered independently and much more tractably.

In this paper, we begin by reviewing the state of the art in aggregate computing and the resiliency properties it currently supports.  
Following a brief discussion of the history of related work in Section~\ref{s:related}, 
we present the notion of computational fields and its elaboration into the aggregate computing ``stack'' of abstractions in Section~\ref{s:stack}, 
and key results on resilience in aggregate computation systems in Section~\ref{s:resilience}.
We then present our view on a roadmap of foundational problems yet to be solved in Section~\ref{s:roadmap}
and conclude with a summary in Section~\ref{s:conclusions}.

\section{History of Related Work}
\label{s:related}

Engineering collective systems has long been a subject of interest in a wide variety of fields, from biology to robotics, networking to high-performance computing, and many more;
a thorough survey of this history may be found in~\cite{SpatialIGI2013}, which we summarise here.
As the foundational issues of engineering collective adaptive systems remain the same, 
particularly when dealing with systems embedded in geometric space and having goals linked to that space (also known as spatial computers), 
a number of common themes have emerged across the multitude of approaches that have been developed.
In particular, there are several clusters of approaches to construct collective adaptive computational behaviours in heterogeneous networks that are identified in~\cite{SpatialIGI2013}:
\begin{itemize}

\item Approaches addressing foundation of group interaction in complex environments mostly extend the archetype process algebra $\pi$-calculus, which models flat compositions of processes, with various versions of environment structure \cite{DBLP:conf/cie/CardelliG10,ambients,Milner200660}, shared-space abstractions \cite{klaim,VCMZ-TAAS2011}, and attribute-based ensembles \cite{SCEL},

\item Device abstraction languages do not provide adaptivity per se, but allow a programmer to focus on adaptivity by making device interaction implicit (e.g., TOTA~\cite{tota}, MPI~\cite{MPI2}, NetLogo~\cite{sklar2007netlogo}, Hood~\cite{hood}),

\item Pattern languages generally provide adaptive means for composing geometric and/or topological constructions, but little computational capability (e.g., Origami Shape Language~\cite{nagpalphd}, Growing Point Language~\cite{coorephd}, ASCAPE~\cite{inchiosa2002overcoming}),

\item Information movement languages are the complement of pattern languages, providing means for summarizing from space-time regions of the environment and streaming these summaries to other regions, but little control over the patterning of that computation (e.g., TinyDB~\cite{tinydb}, Regiment~\cite{regiment}, KQML~\cite{Finin94kqml}),

\item General purpose spatial languages typically require more investment to use as they lack the specialisation of the other categories, but the general constructs they provide avoid the limiting constraints of the other categories (e.g., Protelis~\cite{Protelis15}, Proto~\cite{proto06a}, MGS~\cite{GiavittoMGS02}).
\end{itemize}

Overall, the successes and failures of these language suggest, as observed in~\cite{BPV-COMPUTER2015}, 
that adaptive mechanisms are best arranged to be implicit by default,
that composition of aggregate-level modules and subsystems must be simple, transparent, and result in highly predictable behaviours, 
and that large-scale collective adaptive systems typically require a mixture of coordination mechanisms to be deployed at different places, times, and scales.

\section{Aggregate Programming Approach}
\label{s:stack}

\begin{figure}
\centering
\includegraphics[width=0.5\textwidth]{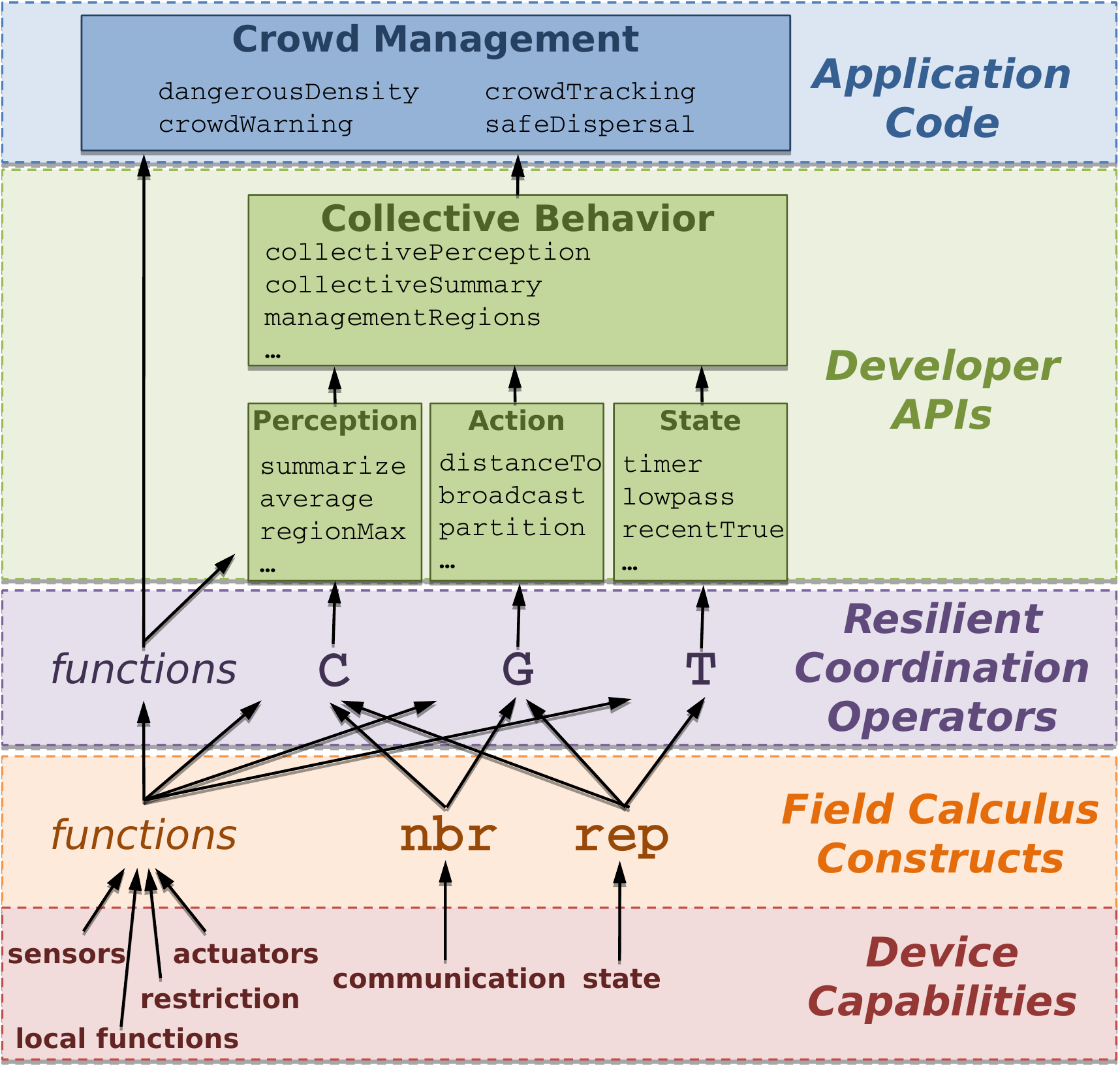}
\caption{Layers of aggregate computing, adapted from \cite{BPV-COMPUTER2015}}\label{f:stack}
\end{figure}

Aggregate computing is an approach that aims to draw on the successes of past approaches to produce a generalised means of programming collective adaptive systems.
Most specifically, the aggregate computing paradigm is grounded in three main concepts: 
\emph{(i)} the reference ``machine'' over which collective adaptive applications run is abstracted to a distributed -- yet conceptually single -- computational device, 
\emph{(ii)} the reference ``elaboration process'' for that machine is the manipulation of a ``collective data-structure'' physically distributed through part or all of the surrounding environment; and 
\emph{(iii)} computation is carried out by cooperation of devices, achieving resiliency by self-organisation.
This approach may then be implemented by the approach of layered abstractions depicted in
Figure \ref{f:stack}, incrementally connecting the capabilities of single devices to the development of collective adaptive applications.
The remainder of this section describes each of these layers in turn.

\subsection{Computational Fields}

A first key question is: what should be the shape of a ``collective data-structure'' manipulated with aggregate computing?
Given the tight connection with physical space for the typical application scenarios we address, we consider collections of values, with each value situated in a specific point of the physical space, and at the time that value has been produced by a device, either directly by a sensor, or as result of some computation.
We define the \emph{domain} of such a collection as a set of space-time points, called \emph{events}; note that such events correspond to computational actions (e.g., sensing, actuation, local computation) executed by some device embedded in the spatial environment.
For any event $\epsilon$ we can thus identify a position $p$ in space, a moment in time $t$, and a device $d$ which computed it (under space and time coordinate system).
%
%For simplicity, we henceforth assume devices be stationary---Section \ref{s:roadmap} will discuss issues in relaxing this hypothesis, and future works.
%
In line with previous works such as \cite{tota,VCMZ-TAAS2011,BealUsbeck12}, we hence rely on the notion of \emph{computational field} (or field for short), defined as a map from a domain (giving the required details of the computational environment) to some set of computational values (e.g., Booleans, numbers, or any complex computational object up to a higher-order description of aggregate behaviour).
%
% I have commented out this line because I believe that this is not true:
%Note that this definition extends that of previous works, that typically consider only space aspects, in a way that allows us to more generally address all the foundational aspects of aggregate computing, especially those related to dynamics of computation.

We sometimes refer to \emph{field evolution} (instead of just a field) to emphasise how computed values evolve over both space and time, and \emph{field snapshot} as a field over a subset of a domain selecting values at a given moment in time---i.e., selecting the last value available in each device at that time.
Computational fields are a very general mechanism, useful to various purposes; one can model inputs coming from the environments as a sensor field, outputs as an actuator field, system knowledge as a data field, any intermediate results of computation as a field of computed values, or the dynamically injected aggregate code to execute as a field of functions.
So, critically, any aggregate computation can be seen as a function from fields to fields, where input and output should have the same domain.

As the shape of a computational field has been clarified, let us consider a second key question: how does our space-time notion of collective data-structure affect the computational model?
In general, computation of a value at a given event $\epsilon$ should depend on some contextual information, certainly including results of computations at the previous event at $\epsilon$'s device and information produced by sensors at $\epsilon$.
Additionally, some notion of local device-to-device interaction is considered.
Embedded in a domain (and depending on application-specific aspects) there is a binary notion of proximity, dictating when two devices are in the neighbouring relation.
It is then assumed that computation of a value at a given event $\epsilon$ can depend on the value at events corresponding to the latest computation at neighbouring devices, that is, assuming a communication that transferred information from a device to its neighbourhood.
Depending on the specific computation to achieve, in particular, neighbouring devices can be restricted to consider only those belonging to a common ``subdomain,'' identifying those devices that cooperatively bring about a common computational goal.

So to recap, computing with fields can be done by leveraging devices' ability to connect with sensors and actuators, to locally compute functions as usual and keep track of results over time, and to communicate with neighbours and possibly do so restricting the proximity relation.
As shown in Figure~\ref{f:stack} (bottom), it is on top of this lowest layer of device mechanisms, and on the notion of field, that aggregate computing grounds and builds higher-levels computing models.

\subsection{Field Calculus}

The field calculus (as expressed in its higher-order version in \cite{HOF-FORTE15}) is the foundation of aggregate computing, as it provides a core language with formalised syntax, semantics and properties, on top of which more accessible programming languages can be built, and resiliency properties can be proven by construction or formal reasoning.
The core idea of field calculus is to express computations by a functional language with the ``everything is a field'' philosophy.
Given an external environment, namely a domain and sensors' values, each expression defines a field on that domain, and function application is a key ingredient that allows one to define reusable behaviour in terms of declaratively-specified transformation from fields to fields; in fact, any field computation takes field evolutions as input  and produces a field evolution as output.
For example, given an input of a Boolean field mapping certain devices
of interest to {\texttt{true}}, an output field of estimated hop-by-hop distances to
the nearest such device can be constructed by iterative
aggregation and spreading of information, such that as the input
changes the output changes to match.
The field calculus succinctly captures the essence of field
computations, much as $\lambda$-calculus~\cite{LambdaCalculus}
does for functional computations or FJ~\cite{FJ} does for object-oriented programming.
A field expression $\e$ is constructed and manipulated using three syntactic program constructs:

\begin{itemize}

\item {\bf Functions:} $\e_\lambda(\e_1,\ldots,\e_n)$ applies the function yielded by
  $\e_\lambda$ to arguments $\e_1,\ldots,\e_n$, with call-by-value semantics.  Such a function can be a
  ``built-in'' primitive (any stateless mathematical, logical, or
  algorithmic function, possibly in infix notation), a sensor or actuator, a function literal ``$(x_1,\ldots,x_n)=>\e$'' or a user-defined function $f$ defined as ``$\textit{def}~f(x_1,\ldots,x_n)\{\e\}$''. 
  For instance, $1+2$ gives a flat field mapping each event to $3$, $\textit{sns-temp()}$ the field of temperatures, \mbox{$((x)=>x+1)(0)$} gives 1 everywhere, and $\textit{mux}(\e_b,\e_1,\e_2)$ computes fields out of $\e_b$, $\e_1$ and $\e_2$, and gives at each event the result given by $\e_1$ where $\e_b$ is \texttt{true} and $\e_2$ where $\e_b$ is \texttt{false}. Importantly, since $\e_\lambda$ is an expression it actually provides a field, namely, a field of functions which could change over space and time: in that case,
  the resulting field is obtained by preventing information flow between events where evaluation of $\e_\lambda$ differ. 
 Thus, for example, $(\textit{mux}(\e_b,+,-))(2,1)$ splits the domain in two subdomains depending on the {\texttt{true}}/{\texttt{false}} evaluation of $\e_b$, and computes $2+1$ in one and $2-1$ in the other.
  
  \item {\bf Dynamics:} $\textit{{rep}}(\e_0)\{\e_\lambda\}$ defines a field holding the evaluation of $\e_0$ initially, and being updated at each event on a device by applying $\e_\lambda$ to the value held at previous event on the same device. 
For instance, $\textit{rep}(0)\{(x)=>x+1\}$ gives a field counting the number of events at each device.
  
  \item {\bf Interaction:} $\textit{nbr}(\e)$ gathers at each event a map from all neighbours to their latest resulting value of computing $\e$. A special set of built-in ``hood'' functions can then be used to summarise such maps back to ordinary expressions. For instance, $\textit{sumHood}(\textit{nbr}\{1\})$ counts the number of neighbours at each event.

\end{itemize}

\noindent An example using the various constructs is the following \texttt{distance} (or \texttt{gradient}) function:%
%\begin{samepage}
\begin{Verbatim}[frame=single,
                  %baselinestretch=,
                  commandchars=\\\{\}]
\km{def} \fn{distance}(source)\{
    \km{rep}(infinity)\{
       (d) => \pr{mux}(source, 0, \pr{minHood}( \pr{nbrRange}() \pr{+} \km{nbr}\{d\}))
    \}  
\}
\end{Verbatim} 
%\end{samepage}
coloring field calculus keywords red, built-in functions green, and
user-defined functions blue.  This code estimates distance \texttt{d} to
devices where \texttt{source} is \texttt{true}: it is initially infinity everywhere, and is computed over time using built-in
selector \texttt{mux} to set sources to $0$ and other devices by the
triangle inequality, taking the minimum value obtained by adding the
distance to each neighbour (as given by sensor $\textit{nbrRange}$) to its estimate of \texttt{d} (obtained by $\textit{nbr}$).

Critically, this aggregate-level model of computation over fields can also be
``compiled'' into an equivalent system of local operations and message
passing actually implementing the field calculus program on a
distributed system~\cite{DVB-SCP2016,HOF-FORTE15}.
In particular, it defines the \emph{computation round} behaviour, framed as a single computable function to be applied at any event. 

%Generally, field calculus sits on top of device mechanisms, and provide a global-level specification language to express collective adaptive behaviour thanks to the combination of the three key constructs it provides (see Figure~\ref{s:stack}).

\subsection{Building Blocks and Libraries}

A key advantage that aggregate programming inherits from managing computational fields functionally (as distinct from other approaches in which this is done either by diffusion/aggregation rules embedded into data items \cite{tota}, or chemical-like rules embedded in ``space'' \cite{VPMS-SAC2012}) is that it intrinsically supports compositionality.
Out of many different algorithms one can express, it is possible to factor out common behaviour into reusable functional components, all of which specify collective adaptive behaviour in terms of field-to-field transformation.
As in all standard functional languages, this methodology results in the creation of complex APIs defining coherent layers of functions, where layers on top depend on layers below, raising the abstraction layer incrementally from basic ingredients to realisations of entire complex application services---see Figure \ref{s:stack} (top).

\begin{figure}[t]
\centering
\subfigure[Operator {\tt G}]{\includegraphics[width=0.24\textwidth]{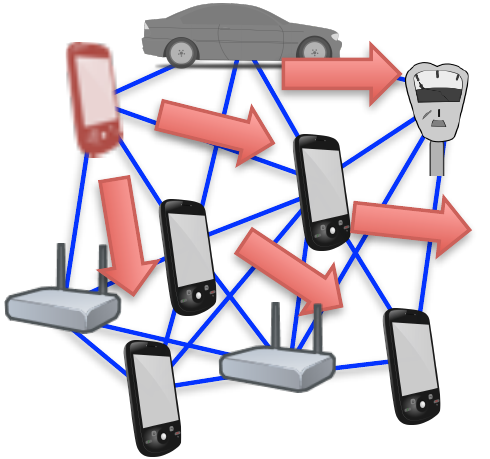}}\hfill
\subfigure[Operator {\tt C}]{\includegraphics[width=0.24\textwidth]{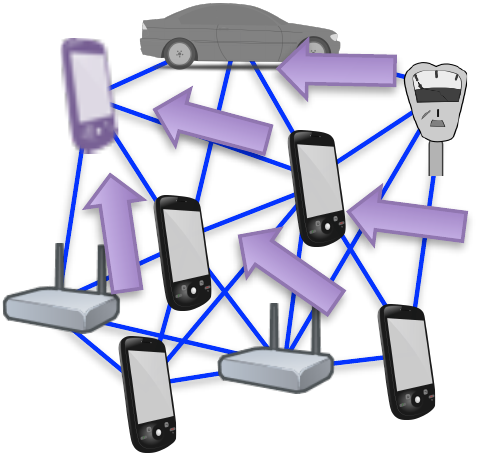}}\hfill
\subfigure[Operator {\tt T}]{\includegraphics[width=0.24\textwidth]{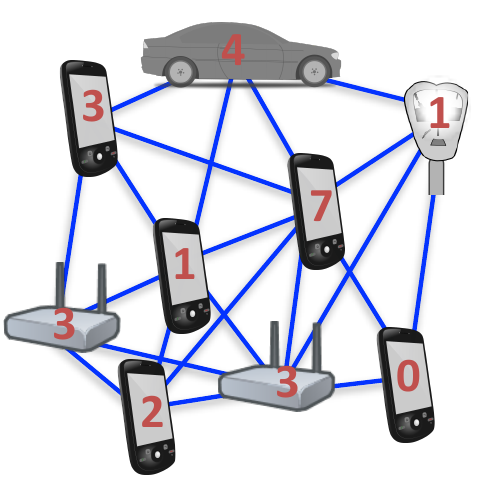}}
\caption{``Building block'' operators for distributed services:
  information-spreading ({\tt G}), information aggregation ({\tt C}),
  and time evolution ({\tt T}), adapted from \cite{BPV-COMPUTER2015}}
\label{f:gct}
\end{figure}

Most notably, experience with programming at the aggregate level and analysis of self-organisation patterns as proposed in literature (see, e.g., \cite{FDMVA-NACO2013}), suggest that the three basic mechanisms one needs in order to ground complex applications include diffusion of information in the network as an advertisement mechanism, aggregation of distributed information as a sensing mechanism, and ``evaporation'' of information as a refresh mechanism.
These three mechanisms can be generally supported by building blocks called \fn{G}, \fn{C} and \fn{T} \cite{BPV-COMPUTER2015}, whose operation is illustrated in Figure~\ref{f:gct} and whose signatures are reported in Figure \ref{f:blocks}.
\begin{figure}[t]
\begin{Verbatim}[frame=single,
                  %baselinestretch=,
                  commandchars=\\\{\}]
\il{Out of source, spread init value, following direction of metric, en-route applying accumulate}
\fn{G}(\vb{source},\vb{init},\vb{metric},\vb{accumulate})

\il{Gathers values of local down potential gradient, en-route accumulating and using null identity value}
\fn{C}(\vb{potential},\vb{accumulate},\vb{local},\vb{null})

\il{Locally apply decay function to initial value, until reaching floor value }
\fn{T}(\vb{initial},\vb{floor},\vb{decay})
\end{Verbatim} 
\caption{Signatures of building blocks}
\label{f:blocks}
\end{figure}
As outlined in \cite{VBDP-SASO15}, different implementations can exist for these building blocks, trading smoothness and speed in different ways.
More importantly, though, a whole set of library functions can be built just on top of \fn{G}, \fn{C} and \fn{T}, by composition of these functions with one another and local functions.
Figure \ref{f:stack} illustrates an example of the various sorts of APIs one can build, up to application services, e.g., used for large crowd management as described in \cite{BPV-COMPUTER2015}.

\section{Results on Resilience}
\label{s:resilience}

Resilience may be generally defined as the ability to adapt to unexpected changes in working conditions.
It is a key property for collective adaptive systems to manifest, used to ensure that system goals can be achieved even in spite of certain classes of change.
Particularly important is the ability of a computing framework to provide a high degree of \emph{inherent} resilience.
This means that the system specification produced at design-time does not explicitly deal with planning or execution of adaptation; rather, it is the underlying framework that is burdened with the goal of dynamically adapt to changes, autonomously finding ways for the system goals to be automatically achieved.
Aggregate computing is a framework able to support inherent resiliency to a rather large set of changes, especially when coupled to specific techniques to structure a system specification.
Technical results on resilience in aggregate computing are reviewed in this section.

\subsection{Resilience to Occasional Disruption: Self-Stabilization}

A typical resilience scenario is the one where a system is in good working condition until a certain occasional event from the environment creates some disruption: at that point, we wish the system to repair itself and eventually return to a good working condition.
For computational fields, this notion has been formalised in \cite{DV-LMCS2015}.
A field expression $\e$ is said to \emph{self-stabilise} if there exists a time $t$ such that, if the environment (position and proximity of devices, and sensor fields) does not change in any $t'>t$, then eventually, at a time $t''>t$, the field resulting from $\e$ will no longer change either, i.e., it reaches a \emph{self-stabilised state} (a field snapshot stable over time).
Additionally, this self-stabilised state must be unique, depending solely on the stable state of the environment and on field expression $\e$, and \emph{not} on the field snapshot at time $t$.
Put in another way, self-stabilisation of a field expression $\e$ implies that, given a stable state of the environment $E$, the resulting field necessarily eventually reaches a stable snapshot $\phi_{\e,E}$ which solely depends on $\e$ and $E$: $\phi_{\e,E}$ can be considered as the result of computation of $\e$ with ``environmental condition'' $E$.
While the distance function given in the previous section enjoys this property, other similar functions are subtly not self-stabilising, like the following gossip function which keeps gossiping the minimum value of an input field across space and time \cite{PBV-COORD2016-LNCS9686}:

\begin{Verbatim}[frame=single,
                  %baselinestretch=,
                  commandchars=\\\{\}]
\km{def} \fn{gossipMin}(field)\{
    \km{rep}(infinity)\{
       (v) => \pr{min}(field, \pr{minHood}( \km{nbr}\{v\}))
    \}  
\}
\end{Verbatim} 
This is not self-stabilising because field evolution never recovers from a change in which the input field (assuming it is taken from a sensor) temporarily flips to a very low value $v$ at some event: even after the value rises back again, $v$ keeps being gossiped in the network.

We here refer to self-stabilisation as a notion of ``resilience to occasional disruption'' because the above definition implies that in a situation of continuous changes in the environment, field evolution keeps chasing a self-stabilised state, but will reach it only if there is enough time following the last change.
Hence, in a situation of continuous changes, self-stabilisation per se provides no guarantee of resilience.

Self-stabilisation is undecidable in general, given that computational rounds are not even guaranteed to terminate due to the universality of local computation.
Thus, ensuring self-stabilisation is a matter of isolating fragments of the calculus that produce only self-stabilising field expressions.
This problem has been addressed in \cite{VBDP-SASO15}, where the following technical results are provided: \emph{(i)} building blocks \fn{G}, \fn{C} and \fn{T} are proved self-stabilising, and \emph{(ii)} by generalising over them, a fragment of the field calculus guaranteeing self-stabilisation is identified.
Most notably, such a fragment is closed under functional composition.
As a result, any library or application built on top of \fn{G}, \fn{C} and \fn{T}, and avoiding direct use of \texttt{rep} (like those showed in Figure \ref{f:stack} and in \cite{BPV-COMPUTER2015}), is self-stabilising by construction.

\subsection{Resilience to Device Distribution: Eventual Consistency}

A weakness of the above property is that the result of computation, expressed as the stabilised field snapshot, may be highly dependent on network shape. 
Even small perturbations to the position of a device, to the proximity relation, or to the addition/removal of a device, can make the field stabilise to a completely different result.
This means that even general aspects like the overall density of devices in a given portion of space can significantly affect the result of computation.
A simple example is given by the following hop-count distance measure, estimating distances only based on the number of hops to a source:

\begin{Verbatim}[frame=single,
                  %baselinestretch=,
                  commandchars=\\\{\}]
\km{def} \fn{hopCountDistance}(source)\{
    \km{rep}(infinity)\{
       (d) => \pr{mux}(source, 0, \pr{minHood}( \km{nbr}\{d\} + 1))
    \}  
\}
\end{Verbatim} 

There, doubling the density of devices while keeping a constant number of neighbours generally results in an increase of hop-count distances.
Since practically the actual location of devices in a pervasive environment can be not known {\it a priori}, and even occasional changes to distribution can be the norm, one may want to introduce more specific forms of self-stabilisation, able to well tolerate changes to device distribution.
Put in another way, we seek a property such that a field expression $\e$ necessarily eventually reaches a stable snapshot $\phi_{\e}$ that depends on $\e$, and is ``mostly independent'' to the shape of the environment, especially at sufficiently high densities.

The work in \cite{BVPD-SASO16} address this issue by a notion of \emph{eventual consistency}, essentially stating that, in addition to self-stabilisation, with the limit of event densities (devices and their work frequency) going to infinity, the stabilised state of computation converges.
This notion of convergence is given by interpreting field snapshots as measurable functions over a continuous domain, and checking whether the Lebesgue integral of the absolute difference between the field snapshot obtained with a given density and that at infinite density actually converges to 0 as density goes to infinity.
Though this notion does not measure the extent to which a device distribution change affects the result of computation, it can give guarantee of robustness to changes in the scale of the number of devices: at sufficient high densities, e.g. a disrupting change like increasing by 1 the order of magnitude of device densities is not going to significantly affect the shape of the stabilised field snapshot.
So, one can easily expect that simpler changes like addition/removal/relocation or one or more devices will likely be irrelevant to the overall computation.

Ensuring eventual consistency is harder than simple self-stabilisation, because of a \emph{boundary} problem.
Many computations involve discrete approximations of non-continuous built-in functions (like test for equality between numbers) which tend to be very fragile to small changes in position (and distances) of devices.
In \cite{BVPD-SASO16}, GPI calculus is introduced as a fragment of field calculus (a fragment significantly smaller than the one of self-stabilisation in \cite{VBDP-SASO15}) which is based on two mechanisms.
First, the only allowed form of field evolution is with a ``Gradient Path Integral'' construct, essentially spreading information outward from a source $s$ and returning at each device $d$ the result of computing the integral of a provided function across the shortest path connecting $s$ with $d$. Examples of fields one can create with this construct include distance measures, broadcasts, and obstacle forecasting, all possibly realised with different kinds of metrics and combined arbitrarily.
Second, expressions that can lead to fragile ``boundary'' values (due to use of non-continuous functions) are marked, such that values cannot differ over any significant region of the field.

\subsection{Resilience to Ongoing Perturbations: Controlling Dynamical Performance}

What kind of resiliency support can we provide in the case of ongoing perturbations of the environment?
There, it is not sufficient simply to know that a system self-stabilises, but it is very important \emph{how} self-stabilisation is reached.
Depending on the application context, we might simply seek fast self-stabilisation, while in other cases we can tolerate slow self-stabilisation provided there is smoothness, i.e., field evolution never shifts to snapshots that are too distant from the actual result of self-stabilisation once reached.
While fast self-stabilisation can be useful with frequent, though non-continuous changes, smooth self-stabilisation may be needed with continuous changes, as in the case of many mobile networks.
Two contributions have been provided in the direction of better controlling field dynamics, so far.

First, in \cite{VBDP-SASO15} an engineering methodology is presented in which \fn{G}, \fn{C} and \fn{T} are selectively replaced with alternative and more specialized implementations that can better trade off speed with adaptiveness in certain contexts of usage.
For instance, the approach in \cite{flexgradient} can be used to compute distances instead of by the standard implementation of \fn{G}, especially when direction of movement to the source is more important than actual estimation of distance, while a multi-path collection of information can be used instead of \fn{C}'s single path one when reactivity to network changes is more important than reactivity to changes in the collected data.

Second, in \cite{PBV-COORD2016-LNCS9686} a technique is proposed to turn gossiping into a self-stabilising process by means of running multiple replicas of gossiping in parallel at staggered times.
If the proper duration of such replicas can be statically estimated, replicated gossip provides a much more controlled evolution of dynamics.
As suggested in next section, this approach might be evaluated as a general meta-technique to improve speed and smoothness of self-stabilisation.

%Third, in [JAKE on CONTROL THEORY PAPER]

\section{Roadmap of Foundational Problems}
\label{s:roadmap}

The results reviewed so far represent important progress in methods for the engineering of collective adaptive systems.
Many foundational questions remain to be addressed, however, and resolving these questions will both broaden the applicability of aggregate computing and improve the guarantees of resilience and performance that can be made.
We now present our view on the critical foundational problems still to be addressed, organising the current key open foundational problems into four thematic groups: universality, static properties, dynamic properties, and workflow constructs.

\subsection{Universality}

The notion of computational universality has long been well-developed both for individual devices and for networks of devices.
In this sense, there is a trivial sense in which aggregate computing can be readily shown as being universal, through the computational universality of the individual devices in the aggregate.
At the aggregate level, however, studying universality helps reasoning in terms of expressiveness, allowing one to understand whether a given choice of language constructs is sufficient to express all required behaviour, and to assess comparison between different languages.

\begin{itemize}
 \item \emph{Discrete notion of universality.} A first notion of universality can be achieved by looking at which kinds of computations one can achieve on a given domain (defined as a finite set of events as of Section 3). Reasonable hypotheses there are that each device can compute with universal Turing power, and that inputs come from values in the local context (sensors and neighbour events). 
 
 \item \emph{Continuous notion of universality.} The work in \cite{BVD-SCW14,bealBasisSCW10} suggests a different notion of universality, that focusses instead on the ability of field computations to generate fields defined over continuous space and time. 
 Similar hypotheses here are that such fields can be locally effectively computed, and that information at an event $\epsilon$ can solely depend on information from the cone of past events from which $\epsilon$ is reachable considering a certain maximum velocity of information. With this notion, field calculus is argued to be universal in \cite{BVD-SCW14}. Considering less specific versions of universality is a key future work.
 
 \item \emph{Consistency between notions universality.} Clearly, many notions of universality can be defined, and hence it will be key to compare and connect them. The work in \cite{VBDP-SASO15} already connects discrete and continuous domains for defining the notion of eventual consistency, which can inspire the definition of a unified notion capturing both discrete and continuous domains.
 
 \item \emph{Mobile devices.} The notions of continuous computation presented in \cite{BVD-SCW14,bealBasisSCW10} address only stationary devices, while in many real-world systems the devices either move themselves or are moved by external forces (e.g., a personal device carried by its owner).  
Consistency between continuous and discrete computational models needs to extend to these cases, as well as accounting for the qualitatively different behavior between tightly packed (``solid''), loosely packed (``liquid''), and sparse (``gas'') distributions of mobile devices.
 \end{itemize}

\subsection{Static Properties}

A key advantage of aggregate computing compared to other approached for designing self-organising systems is its ability to compositionally and declaratively express complex behaviour.
Its functional nature, in particular, allows one to readily reason formally on the expected behaviour of a program.
Many interesting results have already emerged in the area of ``static properties,'' namely, properties of the result of computation, neglecting transitory aspects that concern dynamics of evolution, but there are important areas for which these should be further extended.

\begin{itemize}
 \item \emph{Fragments of resilient behaviour.} As described in previous section, in \cite{VBDP-SASO15} a fragment of self-stabilising field expressions has been identified by generalisation of building blocks \fn{G}, \fn{C} and \fn{T} into specific usage patterns for \textit{rep} construct. Such patterns require to inspect whether certain sub-expressions enjoy properties of monotonicity, boundedness, progressiveness and so on. The work in \cite{DV-LMCS2015} shows how automatically proving such properties in practice is not very easy. Important future work here is to find a larger fragment, with patterns easier to automatically check.
 
 \item \emph{Beyond existing building blocks.} A reason for the current limited extent of the fragment of self-stabilising expressions is due to the fact that it originated from \fn{G}, \fn{C} and \fn{T}, which were identified as reusable blocks even before the self-stabilisation property was established. These three building blocks allow one to functional compose operations of collection and spreading of information, along with functions taking into account timing mechanisms. Although quite expressive, these do not cover all of the useful patterns of self-stabilising algorithms. Identifying new building blocks is key to enlarge the set of resilient aggregate behaviours one can engineer. Areas of future work in this context include but are not limited to graph-based algorithms, adaptive leader election, clustering of data, flocking, and so on.
  
 \item \emph{Model-checking and other formal methods.} 
Recently, formal models are increasingly investgated to predict quantitative and qualitative properties of collective adaptive systems.
To trade off verification time with accuracy, statistical model-checking \cite{stat-mc} is often used instead of classical model-checking in addition to standard simulation, though it only partially alleviates the scalability problem. Recently, fluid flow and mean-field approximations have been proposed to turn large-scale computational systems into systems of differential equations that one could solve analytically or use to derive an evaluation of system behaviour \cite{DBLP:journals/scp/LatellaLM15,DBLP:conf/popl/CardelliTTV16}. We believe that research on aggregate computing can aim at going beyond existing uses of such techniques, more directly addressing space-time considerations and relationship with countinuous notions of computational fields.

 %\item \emph{Beyond current resiliency properties.} As we currently identified the properties of self-stabilisation and eventual consistency to better reason in terms of the result of computations, additional properties are probably needed to extend our ability of engineering various collective adaptive behaviours.
 
\end{itemize}

\subsection{Dynamic Properties}

As discussed in Section 4, the framework of self-stabilisation, though rather expressive, does not address a number of issues of high practical impact, including performance issues as well as quantitative considerations related to transitory errors in the expected behaviour. 
Though rather difficult to address in general, study of dynamic properties is a key ingredient for future research on aggregate computing. 

\begin{itemize}
 \item \emph{Characterisation of resilience.} We believe that a first step towards a more clear understanding of the problem is to analyse the full spectrum of resilience, so as to identify what kind of changes an aggregate system should aim at adapting to, and the extent to which this is done in a proper and satisfactory way.
 
 \item \emph{Speed and smoothness of self-stabilisation and eventual-consistency.} Even when considering self-stabilisation, we find it key to identify formal means by which one can check, control, and then enact, various levels of speed to self-stabilisation, or of smoothness, defined as the ability of evolving towards a stable state along a trajectory guaranteeing good intermediate results. Key issues in this context include finding building block implementations for which extensive empirical analysis can be conducted to study dynamic properties, and addressing the more general problem of how properties of dynamics of certain components are preserved (or at least bounded) by composition. 
 
 \item \emph{Meta-algorithms for resiliency of dynamics.} Of great interest are those techniques that can be applied to a large class of aggregate computations that can improve their resilience, either in terms of turning non-self-stabilising computations into self-stabilising ones, speeding up self-stabilisation, or generally smoothing behaviour. Replicated instances, as initially studied in \cite{PBV-COORD2016-LNCS9686}, are an example of such a technique, which has to be more systematically studied to identify applicability, methodologies for tuning parameters, and extensions to advance flexibility.

\end{itemize}

\subsection{Workflow Constructs}

The functional paradigm adopted by aggregate computing promotes a clear design of the interface of pieces of collective adaptive behaviour, paving the way towards composition, reuse, and substitutability.
On the other hand, simple composition may itself be quickly found too limited in expressive complex interactions between modules. More generally, thinking about aggregate computations in terms of workflow (e.g., sequencing of processes) will be important for dealing with a number of complex real-world applications.

\begin{itemize}
 \item \emph{From fields to processes.} How might we deal with a multiset of interacting processes, as typically considered in process calculi, in the context of aggregate computing? 
 Answering this question is key for a number of important results to be achieved, particularly for defining execution platforms for ecosystems of pervasive computing services. 
 Possibly, this can be addressed by new constructs for the field calculus, capturing parallel composition, interleaved execution, and forms of aggregate interaction. The \texttt{alignedMap} mechanism exploit in \cite{PBV-COORD2016-LNCS9686} is an initial attempt in that direction.
 
 \item \emph{Workflow constructs.} As a notion of process is correctly identified and supported by the field calculus, new building blocks will be needed to expressively compose such processes. It will be needed to clearly identify the distributed starts and ends of a process, so as to support process sequencing, join, fork and similar workflow constructs. Likewise, virtual-machine aspects like handling of exceptions and garbage collection need to be supported in order to provide a full framework for executing complex processes at the aggregate level.
 
\end{itemize}

\section{Conclusions}
\label{s:conclusions}

Aggregate programming is an emerging approach to the engineering of collective adaptive systems.
The layered approach advocated by aggregate computing rests on the core computational model of field-based programming embodied in field calculus.
Resilience is then provided by restriction to building blocks that both provide desired resilience properties and that preserve these properties when composed with one another:
to date, self-stabilization provides resilience to occasional disruptions, eventual consistency provides resilience to distribution of devices, and substitutability can be used to improve the dynamical performance of systems.

Looking toward the future, we have presented a roadmap organising the key foundational problems for advancing aggregate computing.
Beyond this roadmap, there are also a number of pragmatic challenges to address, such as:
improvement of aggregate programming software tools and  language implementations;
characterisation and optimisation of costs in computation, communication, energy consumption, and the like;
extension of the libraries and APIs;
and development of additional tools and other aspects of the engineering ecosystem.
Finally, ongoing work on applications will both realise the value of these approaches into real engineered systems as well as presenting challenges that we expect to feedback into foundational and practical investigation.

%\nocite{*}
\bibliographystyle{eptcs}
\bibliography{long}
\end{document}